\documentclass[vecphys]{svmult}

\usepackage{makeidx}         
\usepackage{graphicx}        
\usepackage{multicol}        
\usepackage[bottom]{footmisc}

\usepackage{amsmath}
\usepackage{epsfig}

\makeindex             

\def\as{\alpha_\mathrm{s}}

\def\PT{\mathrm{PT}}
\def\NP{\mathrm{NP}}
\def\LQCD{\Lambda_\mathrm{QCD}}

\begin{document}

\title*{Resummations in QCD: recent developments}
\author{Andrea Banfi\inst{1}}
\institute{
Universit\`a degli Studi di Milano Bicocca and INFN, Sezione di Milano, Italy
\texttt{andrea.banfi@mib.infn.it}
}
%
%
\maketitle

There are processes which involve two widely separated scales $Q$ and
$Q_0$, where $Q$ is a hard scale, and real emissions characterised by
momentum scales larger than $Q_0$ are not measured, so that only
their virtual counterpart contributes to physical observables. Since
both real and virtual corrections are infrared and collinear (IRC)
divergent, but their sum is finite, incomplete real-virtual
cancellations give rise to large logarithms $L=\ln Q/Q_0$ in
perturbative (PT) expansions, which spoil their convergence.

PT predictions should therefore be improved by resumming logarithmic
enhanced terms to all orders. In the luckiest cases, a resummed cross
section $\sigma(L)$ can be written in the exponentiated form $\sigma(L)=\exp\{L
g_1(\as L)+g_2(\as L)+\as g_3(\as L)+\dots\}$, where $g_1, g_2, g_3$,
etc. contain leading logarithms (LL, $\as^n L^{n+1}$), next-to-leading
logarithms (NLL, $\as^n L^n$) next-to-next-to-leading logarithms
(NNLL, $\as^n L^{n-1}$) and so on.

The basis of resummation is all-order factorisation of IRC
divergences~\cite{CSS}. Any IRC singular contribution to a Feynman
graph ${\cal G}$ (a.k.a. leading region ${\cal G}_L$) can be written
schematically as the following convolution:
\begin{equation}
  \label{eq:GL}
  {\cal G}_L = H \otimes \prod_{\ell=1}^{n_\ell} J_\ell \otimes S\,,
\end{equation}
where:
\begin{enumerate}
\item $H$ is the hard vertex containing lines whose virtuality is of
  order of the hard scale of the process;
\item $n_\ell$ jet functions $J_\ell$, one for each hard massless leg
  $\ell$, containing all collinear singularities;
\item a soft function $S$ embodying infrared singularities.
\end{enumerate}
Since the contribution of ${\cal G}$ to a physical cross section is
obtained by summing over all possible final-state cuts, it is clear
that virtual corrections are observable independent, while the
contribution of real emissions is what discriminates an observable
from the other, and understanding it is the key to resummation.
In particular there is a considerable difference between inclusive
observables, like total cross sections, in which one is not interested
in the structure of the final state, and final-state observables, like
event-shape distributions, where one puts a direct veto on real
emissions.

For inclusive observables, real contributions are factorised via an
integral transform, and the transformed cross section $\tilde\sigma(L)$
exponentiates:
\begin{equation}
  \label{eq:resum-inclusive}
  \tilde\sigma(L) \simeq C(\as)\> e^{E(\as,L)}\,.
\end{equation}
The best known examples of inclusive resummations are
threshold~\cite{threshold} and transverse momentum~\cite{transverse}
resummations.  Here we briefly present results~\cite{higgs-xsct,
  higgs-pt} for Higgs production in hadron-hadron collisions, the
knowledge of which is fundamental for the LHC. Results for other
processes can be found in refs.~\cite{threshold} and~\cite{transverse}.

Threshold resummation is needed for the total Higgs cross section when
the available partonic energy $\sqrt{\hat s}$ is close to the Higgs mass
$M_H$.  In this limit only soft emissions are allowed and the cross
section develops large logarithms $\ln(1\!-\!M_H^2/\hat s)$. After a Mellin
transform these become logarithms of $N$, the variable conjugated to
$M_H^2/{\hat s}$, and at threshold $N\!\to\!\infty$.  The exponent
$E(\as,\ln N)$ is given by
\begin{multline}
  \label{eq:exponent-threshold}
  E(\as(M_H^2),\ln N) = \int_0^1\!\! dz \frac{z^{N-1}-1}{1-z} \times \\
  \times 
  \left[
    2 \int_{M_H^2}^{(1-z)^2 M_H^2}\frac{dq^2}{q^2}A(\as(q^2))+
    D(\as((1-z)^2 M_H^2)) 
  \right]\,.
\end{multline}
The function $A(\as)$ contains soft and collinear contributions
(double logarithms), while single logarithms are embodied in the
function $D(\as)$. Both functions have an expansion in powers of
$\as$.  Our knowledge of $E(\as,\ln N)$ extends at NNLL level, i.e. we
know $A_3$~\cite{A3} and $D_2$~\cite{higgs-xsct}.\footnote{Actually
  also $D_3$ is known~\cite{D3}. However, since $A_4$ is still unknown, we
  cannot push resummations beyond NNLL accuracy.} The numerical results
of ref.~\cite{higgs-xsct} show one of the main benefits of
resummation, the considerable reduction of renormalisation and
factorisation scale dependence.

Large logarithms $\ln (q_T/M_H)$ in the Higgs transverse momentum
distribution $d\sigma/dq_T$ are resummed via a Fourier transform.
In the space of impact parameter $b$ (conjugated to $q_T$) the
resummed exponent is~\cite{higgs-pt} 
\begin{equation}
  \label{eq:exponent-pt}
  E(\as(M_H^2),\ln b) = -\int_{b_0^2/b^2}^{M_H^2}\frac{dq^2}{q^2} 
  \left[A(\as(q^2))\ln\frac{M_H^2}{q^2}+ B(\as(q^2))\right]\,, 
\end{equation}
where $A(\as)$ is the same as for threshold resummations, while single
logarithms build up the function $B(\as)$. The exponent in
eq.~(\ref{eq:exponent-pt}) is known up to NNLL accuracy, i.e. we know
the second order coefficient $B_2$~\cite{higgs-pt}. Not only does
resummation reduce renormalisation and factorisation scale
uncertainties, but also ensures that the resulting $q_T$ spectrum
vanishes linearly at small $q_T$, as is expected on physical grounds.

In recent years a method to combine threshold and transverse momentum
resummation has been developed. This procedure, called joint
resummation, has been exploited to extend transverse momentum
resummations to the threshold region~\cite{joint}, where
eq.~(\ref{eq:exponent-pt}) breaks down, and to improve the description
of the $q_T$ spectra of prompt photons~\cite{joint-prompt} and heavy
quarks~\cite{joint-heavy} .

There are also processes, involving low momentum scales, where a fixed
logarithmic resummation is not sensible, since the factorial
divergence of the PT exponent (always present in QCD) cannot be
neglected.  For instance in the decay $b \!\to\! s\gamma$, after
resummation of all terms responsible for the divergence, the $N$-th
moment $F_N(M_B)$ of the photon energy spectrum can be written as
\begin{equation}
  \label{eq:btosgamma}
  \begin{split}
    F_N(M_B) &= F_N^{\PT}(m_b)\times e^{(N-1)\frac{M_B-m_b}{M_B}}\times 
    F_N^{\NP}\left((N-1)\frac{\LQCD}{M_B}\right)\,,\\
    F_N^{\PT}(m_b) &= \exp\{
    \frac{C_F}{\beta_0}\int_0^\infty \frac{du}{u}T(u)
    \left(\frac{\LQCD}{m_b}\right)^u \times \\
    &\times[{\cal B}_S(u) \Gamma(-2u)(N^{2u}-1)+{\cal B}_J(u)\Gamma(-u)(N^u-1)]
    \}\>.
  \end{split}
\end{equation}
The functions ${\cal B}_J(u)$ and ${\cal B}_S(u)$ are the Borel
transforms of the jet and soft function respectively (see
eq.~(\ref{eq:GL})). The factorial divergence of the PT exponent
reflect in poles away from $u=0$ in the integrand in
eq.~(\ref{eq:btosgamma}). Regulating the $u$-integral with a principal
value prescription and using the pole $b$-quark mass $m_b$, one finds
that all leading power corrections depend only on the mass difference
$M_B\!-\!m_b$ between the meson and the quark, while higher power
corrections, contained in the function $F_N^{\NP}$, turn out to be not
very important, so that one can even attempt a measure of $m_b$ and
$\as$~\cite{btosgamma}.

For final-state observables, such as event-shapes or jet-resolution
parameters (see~\cite{DS-review} for a recent review), a general
statement concerning exponentiation of resummed distributions does not
exist.  This is mainly due to the fact that veto conditions on real
emissions differ from one variable to the other, and in many cases a
full analytical resummation is unfeasible.  However there is a class
of variables $v$, which share the properties of globalness and recursive
infrared-collinear (rIRC) safety~\cite{caesar}, for which one can
resum the corresponding rate $\Sigma(v)$ at NLL level.  More
precisely, one can show that for rIRC safe variables all leading
logarithms (and part of the NL logarithms) exponentiate, and the
remaining NLL contributions factorise:
\begin{equation}
  \label{eq:sigma-resum}
  \Sigma(v) = e^{-R(v)} {\cal F}(R')\,,\qquad
  R'=-v\frac{dR}{dv}\,.
\end{equation}
Both the exponent $R(v)$ and the correction factor ${\cal F}(R')$ can
be computed via a general master formula. This makes it possible to
perform the resummation in a fully automated way,
as implemented in the program \textsc{caesar}~\cite{caesar}.

For other variables, for instance the so-called non-global
variables~\cite{nonglobal1}, the situation is less clear. Non-global
variables measure radiation only in a restricted phase space region, a
typical example being the energy flow away from hard jets. In this
case some approximations for multi-parton matrix elements that are
valid for global rIRC variables do not hold any more, and a
resummation can be performed only numerically and in the large $N_c$
limit~\cite{nonglobal1,nonglobal2}.
Furthermore, large logarithms for these observables come only from
soft emissions at large angles, so that one naively expects 
leading logarithms to be of the form $\as^n L^n$. However there has been
recently a claim that in hadron-hadron collisions super-leading
logarithms $\as^n L^{n+1}$ arise at higher orders~\cite{superleading}.
Properties of these new logarithms have not been fully investigated yet.

\vspace{-.3cm}

%
%
%



\printindex
\end{document}